\documentclass{wlscirep}
\usepackage{protools}

\title{Taming numerical errors in simulations of continuous variable non-Gaussian state preparation}

\author[1, *]{Jan Provazn\'ik}
\author[1]{Radim Filip}
\author[1]{Petr Marek}

\affil[1]{Department of Optics, Palack\'y University, 17. listopadu 1192/12, 771 46 Olomouc, Czech Republic}
\affil[*]{provaznik@optics.upol.cz}

\begin{abstract}\noindent
  Numerical simulation of continuous variable quantum state preparation is a necessary tool for optimization of existing quantum information processing protocols. A powerful instrument for such simulation is the numerical computation in the Fock state representation. It unavoidably uses an approximation of the infinite-dimensional Fock space by finite complex vector spaces implementable with classical digital computers. In this approximation we analyze the accuracy of several currently available methods for computation of the truncated coherent displacement operator. To overcome their limitations we propose an alternative with improved accuracy based on the standard matrix exponential. We then employ the method in analysis of non-Gaussian state preparation scheme based on coherent displacement of a two mode squeezed vacuum with subsequent photon counting measurement. We compare different detection mechanisms, including avalanche photodiodes, their cascades, and photon number resolving detectors in the context of engineering non-linearly squeezed cubic states and construction of qubit-like superpositions between vacuum and single photon states.
\end{abstract}

\begin{document}

\maketitle



\section{Introduction}

Quantum information theory exploits fundamental features of quantum physics to design protocols and algorithms that offer significant improvements over their classical counterparts~\cite{montanaro2016,nielsen2000,zhong2020,obrien2007}. There are several candidate physical systems suitable for these applications, each with distinct advantages. Continuous variable quantum information processing with light offers feasible and fast generation and manipulation of entangled Gaussian quantum states that are at the core of the information protocols~\cite{braunstein2005,weedbrook2012,adesso2014,asavanant2019,larsen2019,chen2014,asavanant2021,larsen2021}. However, truly universal quantum information processing also requires elements of quantum non-Gaussianity~\cite{lloyd1999,gottesman2001,lachman2019,chabaud2020,chabaud2021}. Protocols based on Gaussian states and Gaussian operations are not universal~\cite{lloyd1999} and can be efficiently simulated on a classical device~\cite{mari2012}. 

For continuous variables of light, the non-Gaussianity is commonly introduced by photon number counting detectors, either the most basic on-off detectors capable of discerning presence of light~\cite{pan2012}, or the more advanced detectors truly distinguishing the photon numbers~\cite{lita2008,calkins2013,marsili2013,harder2016,burenkov2017,sperling2017,endo2021,korzh2020,hopker2019,hlousek2019}. Such detectors can be employed for direct conditional implementation of non-Gaussian operations~\cite{dakna1999,zavatta2004,marek2011,ourjoumtsev2006,tipsmark2011,usuga2010,fiurasek2005,marek2018}, or for conditional preparation of non-Gaussian quantum states \cite{ghose2007,yukawa2013,konno2021,tiedau2019,yoshikawa2018,sangouard2010,huang2016,takase2021,ra2019,su2019,pizzimenti2021,gagatsos2019,gagatsos2021}. The latter can be then used as a resource in deterministic implementation of non-Gaussian gates~\cite{gottesman2001,miyata2016,marek2018}. One thing these approaches have in common is the inherent probabilistic nature of measurement that results in several trade-offs between quality of the implemented operation or the prepared quantum state, the rate with which the desired operation succeeds, and the experimental challenges of the photon number resolving detector~\cite{provaznik2020,korzh2020,hopker2019,endo2021,hlousek2019}. For any given set of realistic detectors and any desired task we then need the ability to faithfully simulate the optical circuit to find out the required parameters leading to the optimal performance, or to find out whether the task is even feasible.

However, numerical simulation of simple quantum optical circuits, even though it is often employed in continuous variable quantum information processing~\cite{miatto2020,killoran2019,quesada2019,gupt2019,bromley2020}, is not a straightforward task. It is burdened by various difficulties, including discretization errors in numerical models relying on continuous representation, truncation errors in discrete models~\cite{miatto2020}, the omnipresent rounding errors due to finite precision of arithmetics~\cite{fox1971,goldberg1991,higham2002,dahlquist2003,heath2018} and numerical truncation errors occurring in finite approximations of infinite processes~\cite{higham2002,heath2018}. If not prevented by rigorous analysis, these numerical artifacts can dominate the computed values and lead to rapid divergence from correct results.

In this paper we evaluate the numerical errors arising when an optical circuit for probabilistic preparation of non-Gaussian quantum states of light \cite{gottesman2001,ghose2007} is simulated on a classical digital computer. We then propose an alternative method for construction of truncated unitary operators aiming to curtail these errors. Finally we take advantage of these tools to fully simulate the circuit for preparation of resource states for the cubic phase gate \cite{miyata2016}, and single mode qubit-like superpositions of zero and one photon. The goal is to find the optimal trade-offs between the quality of the states and the probability of success for a range of available photon counting detectors~\cite{lita2008,provaznik2020,hopker2019,endo2021,korzh2020,hlousek2019}.

This paper is structured as follows. In the second section we review the state preparation circuit. In the third section we describe the errors naturally occurring in numerical simulations. In the fourth section we focus on coherent displacement and identify the numerical errors appearing in different methods of its calculation. In the fifth section we propose an alternative method for its calculation, followed by an overview of verification process in the sixth section. We then proceed with the seventh section, where we describe the methodology of the actual simulation and present the results of its applications in sections eight and nine.


\section{State preparation circuit}

\begin{figure}[h]
  \begin{center}
    \includegraphics[width = 0.50 \columnwidth]{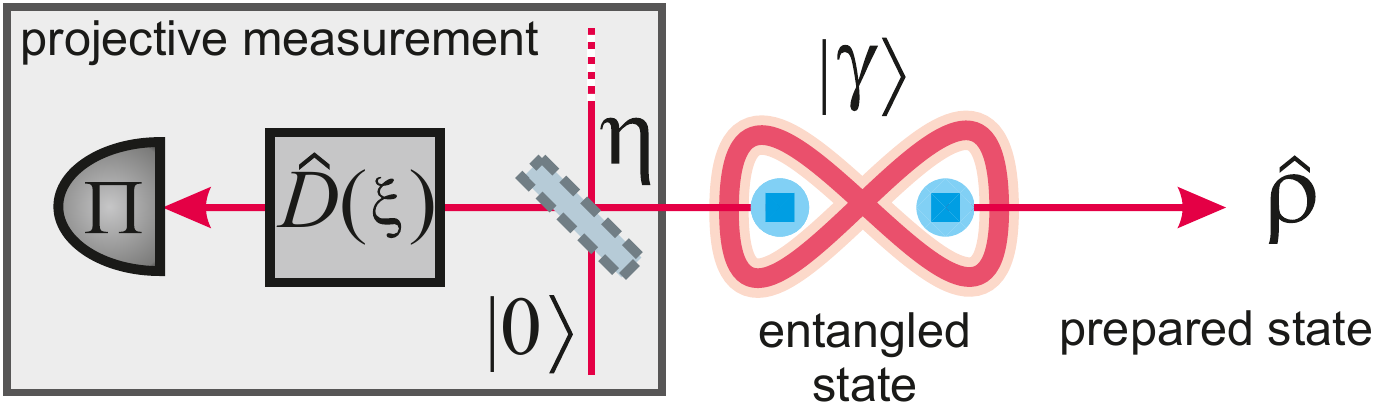}
  \end{center}
  \caption{
    Variation of the conditional preparation scheme. We start with a two mode squeezed vacuum state $\ket{\gamma}$. One of its modes is then displaced with $\hat{D}(\xi)$ and measured, using either APD, PNRD or an APD cascade approximating PNRD. The detection outcome is characterized by the POVM element $\hat{\Pi}$. We model overall losses and inefficiencies within the scheme using a beam splitter with intensity transmittance~$\eta$ to represent attenuation of the signal state in the setup.
  }
  \label{scheme}
\end{figure}

The most common method of conditional state preparation is based on suitable manipulation of EPR state with coherent displacement and subsequent photon counting measurement~\cite{gottesman2001,yukawa2013,yukawa2013b,bohmann2018}. In Fig.~\ref{scheme} we present a variant of the circuit which can be used for preparation of simple non-Gaussian quantum states, including the qubit-like $\ket{0}$ and $\ket{1}$ superpositions. Our circuit accounts for basic imperfections limited to detection inefficiencies and propagation losses. A physical EPR resource, generating the two mode squeezed vacuum~(TMSV), lies at its very heart and serves as a source of perfectly correlated photons. One of the entangled modes is then displaced with controllable amplitude and phase and consequently measured. The detection can use either an avalanche photodiode~(APD), a photon number resolving detector~(PNRD) or its approximation employing an APD cascade \cite{provaznik2020}.
The resulting marginal state
\begin{equation}\label{schR}
  \hatrho =
  P^{-1}
  \trx_{2} \big\{
    \hat{D}_{2} (\xi)
    \mathcal{G}_{2}^{\eta}
    ( \ketbra{\gamma}[1, 2]{\gamma} )
    \hat{D}_{2}^{\dagger} (\xi)
    \hat{\Pi}_{2}(\pi) ]
  \big\} \Qc
\end{equation}
conditioned on the detection outcome $\pi$, characterized by the POVM element $\hat{\Pi}_{2}(\pi)$, is obtained with the probability of success
\begin{equation}\label{schP}
  P = 
  \trx_{1, 2} \big\{
    \hat{D}_{2} (\xi)
    \mathcal{G}_{2}^{\eta}
    ( \ketbra{\gamma}[1, 2]{\gamma} )
    \hat{D}_{2}^{\dagger} (\xi)
    \hat{\Pi}_{2}(\pi) ]
  \big\} \Qd
\end{equation}
In both the expressions \eqref{schR} and \eqref{schP} we use the lower right indices to emphasize which modes the operators and channels act on. Starting from the inner-most component, the initial TMSV state is denoted with ${ \ket{\gamma}_{1,2} = \sum_{i = 0}^{\infty} \mu_{i} (\gamma) \ket{i}_{1} \ket{i}_{2} }$ with coefficients ${\mu_{i} (\gamma) = \cosh^{-1} \gamma \tanh^{i} \gamma}$, where the parameter~${\gamma \in \reals}$ sets the experimentally controllable squeezing strength. We model the overall losses and inefficiencies in the preparation scheme as attenuation of the measured mode prior to its displacement. This can represented by a Gaussian quantum channel $\mathcal{G}_{2}^{\eta} (\hatrho)$ with its action on the mode given in terms of Kraus operators~\cite{ivan2011} as ${ \mathcal{G}_{2}^{\eta} (\hatrho) = \sum_{i = 0}^{\infty} \hat{M}_{2}(i) \hatrho \hat{M}_{2}^{\dagger} (i) }$ with ${ \hat{M}_{2} (i) = \frac{1}{\sqrt{i!}} (\sqrt{1 - \eta})^{i} \sqrt{\eta}^{\hat{N}_{2}} \hat{A}_{2}^{i} }$, where ${\hat{N}_{2} \coloneqq \hat{A}_{2}^{\dagger} \hat{A}_{2}}$ defines the photon number operator and $\hat{A}_{2}$ denotes the annihilation operator respective to the measured mode. The parameter~${\eta \in [0, 1]}$ describes the efficiency of the preparation circuit. Subsequently the converse~${1 - \eta}$ characterizes the overall losses and inefficiencies in the preparation scheme. The displacement of the second mode is given by the unitary operator ${ \hat{D}_{2} (\xi) = \exp( \xi \hat{A}_{2}^{\dagger} - \conj{\xi} \hat{A}_{2}) }$, where ${\xi \in \complexes}$ is the displacement amplitude~\cite{cahill1969}.
In a more realistic analysis of the preparation circuit it would be straightforward to include the propagation losses affecting the mode carrying the resulting state. This form of decoherence can be accounted for by modifying the squeezing strength of the non-linearly squeezed state~\cite{kala2021}. Consequently we do not consider this additional attenuation since it does not influence the fundamental properties of these non-Gaussian states.

From the experimental perspective the parameters $\gamma$ and $\xi$ can be fine tuned to engineer a desired state $\hatrho$ with optimal performance given particular experimental configuration characterized by the efficiency $\eta$ and conditioning on the detection outcome $\pi$ with respective POVM element~$\hat{\Pi}(\pi)$.

We can utilize this scheme to prepare a variety of quantum states. Consider now a lossless configuration employing an ideal PNRD. Its POVM elements correspond to projectors $\ketbra{f}{f}$ onto individual Fock states $\ket{f}$. The output state, conditioned on the detection of a particular Fock state~$\ket{f}$, is then proportional to ${ \sum_{i = 0}^{\infty} \mu_{i} (\gamma) [D(\xi)]_{fi} \ket{i} }$ where the coefficients $\mu_{i} (\gamma)$ follow from the definition of the TMSV state and ${[D(\xi)]_{fi} = \braket{f|\hat{D}(\xi)|i}}$ are matrix elements of the displacement operator. By tuning the parameters $\gamma$ and $\xi$ we can construct a set of states parametrized by the possible combinations of the $\mu_{i}$ and $[D(\xi)]_{fi}$ coefficients.

Possible applications of this scheme include construction of generally non-classical superpositions of Fock states~\cite{ghose2007,yukawa2013} and, in particular, non-linearly squeezed non-Gaussian states~\cite{gottesman2001,yukawa2013b}. Every application can be translated into constrained optimization of the tunable parameters with the constraint and objective functions embodying the nature of the particular application.

For example, if one were to construct a specific state $\ket{\psi}$, the optimization objective could be to maximize some metric of similarity with the target state, e.g., fidelity. It would also be practical to construct the state with non-negligible probability of success. This requirement could be expressed either as a constraint $P \geq \tau$ allowing only solutions with the probability greater than some threshold, or as an additional optimization objective in multi-objective optimization.

The constraint and objective functions generally involve the success probability \eqref{schP} and the resulting density operator \eqref{schR}. Both of which can be obtained by simulating the preparation procedure numerically on a classical digital computer. But alas, numerical simulations come with their own hurdles which will be identified and subsequently addressed in the following sections.


\section{Perils of numerical simulation of CV systems}

Classical digital computers~\cite{turing1937} encode information into finite sequences of bits and it is therefore impossible to represent arbitrary real numbers. The standard approach~\cite{ieee754,goldberg1991,higham2002,dahlquist2003,heath2018} is to approximate real numbers with \textbf{floating point} (FP) numbers. Real numbers are then rounded to their closest representable FP neighbors. This generally introduces \textbf{rounding errors}. To make matters worse, FP arithmetic with FP numbers does not necessarily produce exactly representable floating point numbers. Results of FP arithmetic must be rounded, possibly introducing additional rounding errors~\cite{ieee754,goldberg1991,higham2002,dahlquist2003,heath2018}. Consequently complex sequences of arithmetic operations possess the potential to accumulate and even amplify rounding errors. Even the most straightforward tasks such as adding up a sequence of FP numbers can produce widely different results with varying degrees of accuracy based on the algorithm of choice~\cite{goldberg1991,higham2002,dahlquist2003}. Rounding error analysis is therefore a crucial part of algorithm design~\cite{higham2002,dahlquist2003,goldberg1991,fox1971} and commonly used numerical algorithms are frequently accompanied by rigorous rounding error analysis. Nevertheless numerical simulation cannot be considered completely accurate as the error analysis only establishes upper bounds on the numerical errors~\cite{higham2002,dahlquist2003,heath2018,fox1971}.


The practical concerns, when dealing with numerical simulation, are therefore always related to size of the errors, rather than to their presence. This is a familiar concept in physics, a discipline which is well acquainted with limited precision of measured quantities~\cite{barlow1989,bevington2003}. Numerical simulation of CV systems suffers from further issues related to the fundamental representation of quantum states and quantum operations. CV states reside in infinite-dimensional Hilbert spaces and can be, in principle, described in two distinct ways. The first description employs continuous functions, either wave functions given in position or momentum representation, or quasi-probability distributions~\cite{braunstein2005,weedbrook2012,adesso2014} which combine the two quadratures. The practical issue with this approach is the continuous nature and generally infinite support of these functions as their support must be limited to finite intervals and both their domains and ranges discretized during numerical integration~\cite{dahlquist2003,heath2018}, introducing additional numerical errors.

Alternatively we can take the advantage of the discrete Fock basis spanned by eigenstates of the number operator. This basis is still infinite but, unlike in the case of basis spanned by eigenstates of continuous operators, the number of its elements is \textit{countable}. While exact representation of CV states in Fock basis remains impossible, we can truncate the basis to a finite number of elements and approximate the original Hilbert space with this truncated, finite-dimensional, restriction. We can thusly avoid discretization errors and deal with truncation errors instead. Consequently numerical simulations utilizing truncated Hilbert spaces spanned by truncated Fock basis are often employed in detailed analysis of CV quantum circuits.


\subsection*{Formal definition of truncated Fock spaces}

Let $\sI$ denote the original Hilbert space and let ${S \coloneqq \{ \ket{j} \in \sI \|\; j = 0, 1, \dotsc \}}$ be the original Fock basis (FB). In this basis the vector components of individual Fock states $\ket{j} \in S$ satisfy ${[ \ket{j} ]_{i}^{S} \coloneqq \braket{i|j} \equiv \delta_{ij}}$, that is, Fock states form an orthonormal basis. We take the first $\ff$ elements of FB, $\{ \ket{0}, \dotsc, \ket{\ff - 1} \} \subset S$ and truncate their vector forms to the first $\ff$ components, forming the \textbf{truncated Fock basis} (TFB) ${ S_{\ff} = \{ \ket{0}^{(\ff)}, \dotsc, \ket{\ff - 1}^{(\ff - 1)} \} }$ where we use the upper right indices in $\ket{j}^{(\ff)}$ to denote dimensions of said vectors. Vector components of TFB elements satisfy ${ [ \ket{j}^{(\ff)} ]_{i}^{S_{\ff}} \coloneqq {}^{(\ff)}\negmedspace\braket{j | i}^{(\ff)} \equiv [ \ket{j} ]_{i}^{S} \equiv \delta_{ij} \; \forall i = 0, \dotsc, \ff - 1 }$. The basis therefore remains orthonormal. The linear hull of $S_{\ff}$ forms the $\ff$ dimensional \textbf{truncated Fock space} (TFS)~$\sN$.

So far we have only defined TFS itself and the transition from FB to TFB. In the following we define the transition of vectors from $\sI$ into $\sN$ and linear operators from $\sL(\sI)$ to $\sL(\sN)$. Let $\ket{\psi} \in \sI$ be an arbitrary state expressed as ${ \ket{\psi} = \sum_{i = 0}^{\infty} c_{\psi}(i) \ket{i} }$ (where $\ket{i} \in S$) with coefficients $c_{\psi}(i) = [\ket{\psi}]^{S}_{i} \coloneqq \braket{i|\psi} \in \complexes$. The expression $\trunc_{\ff} \{ \ket{\psi} \} \coloneqq \sum_{i = 0}^{\ff - 1} c_{\psi}(i) \ket{i}^{(\ff)}$ (where ${\ket{i}^{(\ff)} \in S_{\ff}}$) then defines its truncated variant from~$\sN$. Let ${\hat{G} \in \sL(\sI)}$ be a linear operator on $\sI$ expressed as ${ \hat{G} = \sum_{i = 0}^{\infty} \sum_{j = 0}^{\infty} g(i, j) \ketbra{i}{j} }$ (where ${\ket{i}, \ket{j} \in S}$) with matrix elements ${ g(i, j) = [\hat{G}]^{S}_{ij} \coloneqq \braket{i | \hat{G} | j} \in \complexes }$. Then $\trunc_{\ff} \{ \hat{G} \} \coloneqq \sum_{i = 0}^{\ff - 1} \sum_{j = 0}^{\ff - 1} g(i, j) \ket{i}^{(\ff)}\negmedspace\bra{j}$ (where ${\ket{i}^{(\ff)}\!, \ket{j}^{(\ff)} \in S_{\ff}}$) defines its truncated analogue on~$\sL(\sN)$. A natural extension of this approach allows for transitions from higher-dimensional spaces to lower-dimensional spaces.


\subsection*{Navigating truncated Fock spaces}

In this description, pure quantum states become complex $\ff$ dimensional vectors of numbers, linear operators turn into complex $\ff \times \ff$ matrices and the operations we would otherwise perform, reduce to linear algebraic expressions such as matrix multiplication, Kronecker products and matrix traces. There is, however, a hefty price to be paid for this simplification, manifesting in the form of truncation errors with several distinct effects on the simulation.

%

Firstly, it is impossible to represent general quantum states exactly. Take an arbitrary quantum state $\ket{\zeta} \in \sI$ and its truncated variant $\trunc_{\ff} \{ \ket{\zeta} \} \in \sN$. The quality of the truncated state can be determined from its normalization, or rather the lack of it, using the cutoff error
\begin{equation}\label{error}
  {\textstyle \truncerror_{F}} \{ \ket{\zeta} \} \coloneqq
  1 - \sum_{i = 0}^{\ff - 1} \abs{ c_{\zeta}(i) }^{2} \Qc
\end{equation}
where $c_{\zeta}(i) = [ \ket{\zeta} ]_{i}^{S} \equiv \braket{i | \zeta}$ are the vector components of the state $\ket{\zeta}$ in Fock representation. In essence the quality of the representation is loosely given by the support of the state relative to the dimension of the TFS. This is not the only conceivable metric, but it is a convenient one as it is straightforward to calculate.

%

Secondly, the algebraic structure of the space changes with the transition to finite dimension. As a result the usual commutation rules no longer apply since for any pair of operators $\hat{G}$ and $\hat{H}$ the relation ${ \trunc_{\ff} \{ [\hat{G}, \hat{H} ] \} = [ \trunc_{\ff} \{ \hat{G} \}, \trunc_{\ff} \{ \hat{H} \} ] }$ does not necessarily hold anymore. We can illustrate the change in algebraic structure on bosonic creation and annihilation operators. In the regular infinite-dimensional case we have ${[ \hat{A}, \hat{A}^{\dagger} ] = \hat{\eye}}$, that is, the two operators commute to identity. With the truncated commutator the result remains the same ${ \trunc_{\ff} \{ [ \hat{A}, \hat{A}^{\dagger} ] \} = \trunc_{\ff} \{ \hat{\eye} \} \equiv \eye^{(\ff)} }$, which is an identity matrix of the corresponding dimension $\ff$. Conversely the commutator of the truncated annihilation and creation operators differs from identity in the final element on the diagonal
\begin{equation}\textstyle
  { [ \trunc_{\ff} \{ \hat{A} \}, \trunc_{\ff} \{ \hat{A}^{\dagger} \} ] = \eye^{(\ff)} - \ff \ket{\ff - 1}^{(\ff)}\negmedspace\bra{\ff - 1} }
\end{equation}
which can be understood as a truncation error due to the product of two truncated matrices. 


%

Thirdly and finally, replacing infinite-dimensional operators in arguments of operator functions with their truncated versions may not be without consequences. Consider an operator function $f (\hat{Q})$. In principle $\trunc_{\ff} \{ f(\hat{Q}) \} \neq f(\trunc_{\ff} \{ \hat{Q} \} )$ for general operator arguments. This has grave consequences for numerical simulation of unitary evolution. It is customary to approximate the exponential operator, $\trunc_{\ff} \{ \exp(\hat{Q}) \}$, with the matrix exponential $\expm( \trunc_{\ff} \{ \hat{Q} \} )$ of the truncated operator argument~\cite{miatto2020}. However, this method can not be relied upon as ${\trunc_{\ff} \{ \exp(\hat{Q}) \} \neq \expm( \trunc_{\ff} \{ \hat{Q} \} )}$. We must therefore seek alternative approaches: there are three primary techniques available for numerical simulation. The first one relies on the knowledge of a closed form formula for elements of the unitary operator. It has to be derived analytically and is not always attainable. The second method, proposed in the recent paper \cite{miatto2020}, is numerical and derives individual elements of unitaries by recurrent formulae. In the third approach the matrix exponential is simply computed with the truncated matrix argument as $\expm(\trunc_{\ff} \{ \hat{Q} \} )$ and the dimension of the computation space is chosen large enough so that the errors are irrelevant in the particular simulation.

Neither approach is perfect. Each suffers from specific numerical errors. This is a valid concern even for the first method which uses analytical forms: it is because mathematical expressions, especially those involving factorials, large powers of non-negligible numbers or relying on special functions, which are often defined using similar expressions or recurrent formulae, still need to be evaluated numerically with finite precision in floating point arithmetic, leading to introduction and eventual accumulation of rounding errors. The numerical errors cannot be straightforwardly calculated without a priori knowledge of the ideal operator or without thorough numerical analysis of rounding errors, an area of expertise that is mostly out of the scope of theoretical physics and therefore scarcely present in research reports.


In the following section we apply these methods of construction to the simplest experimentally testable example, coherent displacement, and use this particular case study to demonstrate the fundamental shortcomings of each approach.


\section{The curious case of coherent displacement}

Coherent displacement is a fundamental Gaussian operation in quantum optics used in a broad range of quantum protocols for quantum state preparation, manipulation, and measurement~\cite{adesso2014,braunstein2005,gottesman2001,cahill1969,weedbrook2012,lloyd1999}. Coherent displacement is represented by the unitary operator
\begin{equation}\label{dxi}
  \hat{D} (\xi) \coloneqq \exp( \xi \hat{A}^{\dagger} - \conj{\xi} \hat{A})
\end{equation}
where $\xi \in \complexes$ gives the displacement amplitude and $\hat{A}, \hat{A}^{\dagger}$ represent the annihilation and creation operators. It is one of the operations for which a closed form formula exists \cite{cahill1969}, given as
\begin{equation}\label{cahill1969}
  \braket{ m | \hat{D}(\xi) | n }
  =
  \sqrt{\frac{n!}{m!}}
  \xi^{m - n}
  \exp \left(- \frac{1}{2} |\xi|^{2} \right)
  L_{n}^{(m - n)} (|\xi|^{2})
  \Qc
  \quad m \geq n
\end{equation}
where $L_{\beta}^{\alpha} (x)$ denotes the associated Laguerre polynomial function~\cite{bateman1981}. This relation only covers the lower triangular matrix; the rest of the matrix can be easily recovered from~\eqref{cahill1969} using
\begin{equation}
  \braket{m | \hat{D} (\xi) | n} 
  = 
  (- 1)^{m - n} \;
  \conj{( \braket{n | \hat{D} (\xi) |m} )}
  \Qc
  \quad m < n \Qd
\end{equation}
The formula \eqref{cahill1969} can be computed in multiple different ways with varying numerical accuracy impacted by the simplifications made in the expression and the order of their evaluation. When implemented exactly as it stands in \eqref{cahill1969}, it is plagued by the limitations of FP arithmetic. Its first term underflows for comparatively large $m$, while the second term overflows for $\abs{\xi} > 1$ and large enough difference $m - n$. When both the numerical underflow and the overflow coincide, the ill-defined expression $0 \times \infty$ is evaluated, resulting in error. We discuss the circumstances in detail in Section S1 of the Supplementary material and establish a set of acceptable combinations of the $m$, $n$ and $\abs{\xi}$ parameters such that formula \eqref{cahill1969} is always well defined.

We can utilize the recurrent formulae~\cite{miatto2020} or the plain matrix exponential~\cite{higham2005,almohy2010} with a truncated argument instead of the closed form formula~\eqref{cahill1969}. While we can not ascertain their accuracy without a priori knowledge of the ideal operator, we can easily determine whether the generated matrices $G$ are outright incorrect by checking the normalisation
\begin{equation}\label{dxi-norm}
  \norm{ G \ket{j}^{(\ff)} }_{2} 
  \coloneqq \sqrt{ \sum_{i = 0}^{\ff - 1}
      \Abs{ [ G ]_{ij}^{S_{\ff}} }^{2}
    }
\end{equation}
of displaced truncated Fock states $\{ \ket{0}^{(\ff)}, \dotsc, \ket{\ff - 1}^{(\ff)} \}$. It corresponds to the sum of squared absolute values of elements in the $j$th column of the truncated displacement matrix $G \coloneqq \trunc_{\ff} \{ \hat{D} (\xi) \}$ or its approximation employing the matrix exponential $\expm( \trunc_{\ff} \{ \hat{Q} \} )$ with truncated argument where we set ${\trunc_{\ff} \{ \hat{Q} \} \coloneqq \xi \trunc_{\ff} \{ \hat{A}^{\dagger} \} - \conj{\xi} \trunc_{\ff} \{ \hat{A} \} } $.

\begin{figure}[h]
  \includegraphics[width = \columnwidth]{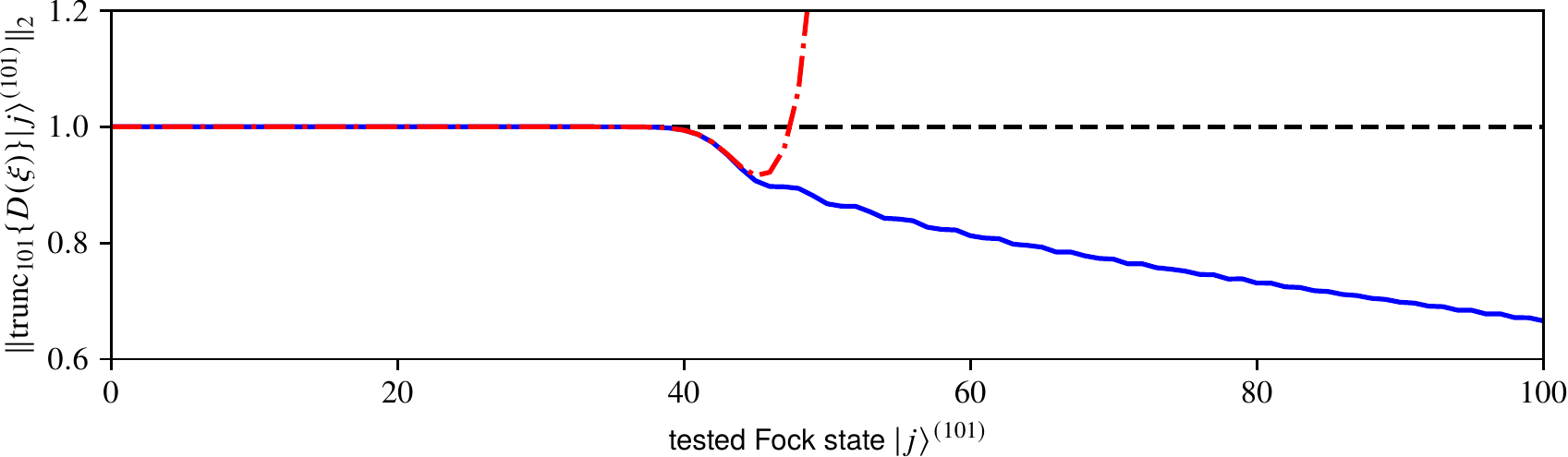}
  \caption{
    Normalisation \eqref{dxi-norm} of individual displaced truncated Fock states $\ket{j}^{(101)}$ with ${0 \leq j \leq 100}$. The displacement operator $\trunc_{101} \{ \hat{D} (3 - 2\imath) \}$ is constructed on $101$ dimensional TFS using the closed form formula (blue solid), the recurrent formula (red dash-dotted), and approximated with the matrix exponential (black dashed line).
  }
  \label{cond2}
\end{figure}

In Fig.~\ref{cond2} we show the normalisation \eqref{dxi-norm} for~${\trunc_{101}\{\hat{D}(3 - 2\imath)\}}$ constructed using the closed form formula~\eqref{cahill1969}, represented with a blue solid line, the recurrent formula~\cite{miatto2020} shown with a red dash-dotted line, and approximated with the matrix exponential (black dashed line). We utilize double precision~\cite{higham2002,ieee754} in the computation and try to avoid numerical issues plaguing the direct method \eqref{cahill1969} by keeping the working dimension sufficiently low. There are two regions of qualitatively distinct behaviour in the plot. The first region, spanning the first $40$ Fock states, shows correct normalization for all three methods of construction. In the following region the normalisation dwindles for both the closed form and the recurrent formulae whilst the matrix exponential remains \textit{incorrectly} normalized. It remains normalized only because the matrix exponential function, by definition, produces unitary matrices from anti-Hermitian arguments. Unitarity is not necessarily the desired outcome here since the goal is to obtain the correct $\trunc_{101}\{ \expm(\hat{Q}) \}$ matrix rather than the computed approximation~$\expm( \trunc_{101}\{ \hat{Q} \} )$.

Let us explicitly discuss the issue at hand. The displacement operator \eqref{dxi} is unitary by definition. Columns of its matrix representation can be understood as coefficient vectors of displaced Fock states. In the infinite-dimensional case these states should be normalized, that is the vector $2$--norm~\cite{golub2013} of each column should satisfy ${ \norm{ \hat{D} (\xi)\ket{j} }_{2} \equiv 1 \,\forall \ket{j} \in S }$. However, this will not generally hold in finite dimension where we can find a threshold state ${\ket{\tau}^{(\ff)} \in S_{\ff}}$ that, when displaced, will not be properly represented on the TFS. The states $j \geq \tau$ will suffer from non-negligible errors~\eqref{error}, making their normalization $\norm{ \trunc_{\ff} \{ \hat{D} (\xi) \} \ket{j}^{(\ff)} }_{2} < 1$.

The plot in Fig.~\ref{cond2} reveals that when the matrix is constructed via~\eqref{cahill1969}, the higher states are correctly denormalized. Conversely the matrix exponential produces \textit{incorrectly} normalized states. In this context such behavior can be considered a manifestation of truncation errors.

The normalisation of the recurrently computed matrix starts to rise exponentially somewhere around~$j \approx 50$ due to accumulation of rounding errors. This behavior depends on the chosen $\xi$ and the breakdown is more prominent when $\xi$ is large. Here the displacement~${\xi = 3 - 2\imath}$ was chosen to emphasize this effect. For instance, when~$\xi = 1$, a similar exponential breakdown appears for~${j \approx 400}$ instead.


\section{Truncated approximate matrix exponential (TAME)}

So far we have seen that, when it comes to numerically generating truncated representations of unitary operators, both direct calculation and the recurrent formulae have fundamental issues leading to significant rounding errors or numerically invalid expressions. The matrix exponential function avoids these issues mostly at the cost of truncation errors and their subsequent amplification. However, the observations in Fig.~\ref{cond2} also suggest that these errors tend to be significant only in higher regions of said matrices.

This opens up a new possibility of approximating the exponential operators. We can use the matrix exponential on a sufficiently higher dimension $d_1$ and only then truncate the result to the required $d_0$, thus avoiding the erroneous areas, while, at the same time, keeping the computational dimension $d_{1}$ low enough to avoid needlessly increasing the time of computation. We call this approach \textbf{truncated approximate matrix exponential} (TAME). Consider the approximation of the truncated displacement operator, $\trunc_{d_{0}} \{ D(\xi) \} \in \sL(\sH_{d_{0}})$, constructed in such a way,
\begin{equation}\label{tame}
  \trunc_{d_{0}} \{ \hat{D} (\xi) \}
  \approx \tame(\hat{Q}, d_{1}, d_{0})
  \coloneqq \trunc\limits_{d_{0}} \Big\{ 
    \expm \big( \trunc_{d_{1}} \{ \hat{Q} \} \big)
  \Big\} \Qd
\end{equation}
Here $d_{1}$ represents the initial working dimension and $d_{0}$ the final dimension of the target TFS. Following~\eqref{dxi} we set ${ \trunc_{d_{1}} \{ \hat{Q} \} \coloneqq \xi \trunc_{d_{1}} \{ \hat{A}^{\dagger} \} - \conj{\xi} \trunc_{d_{1}} \{ \hat{A} \} }$.

\begin{figure}[h]
  \includegraphics[width = \columnwidth]{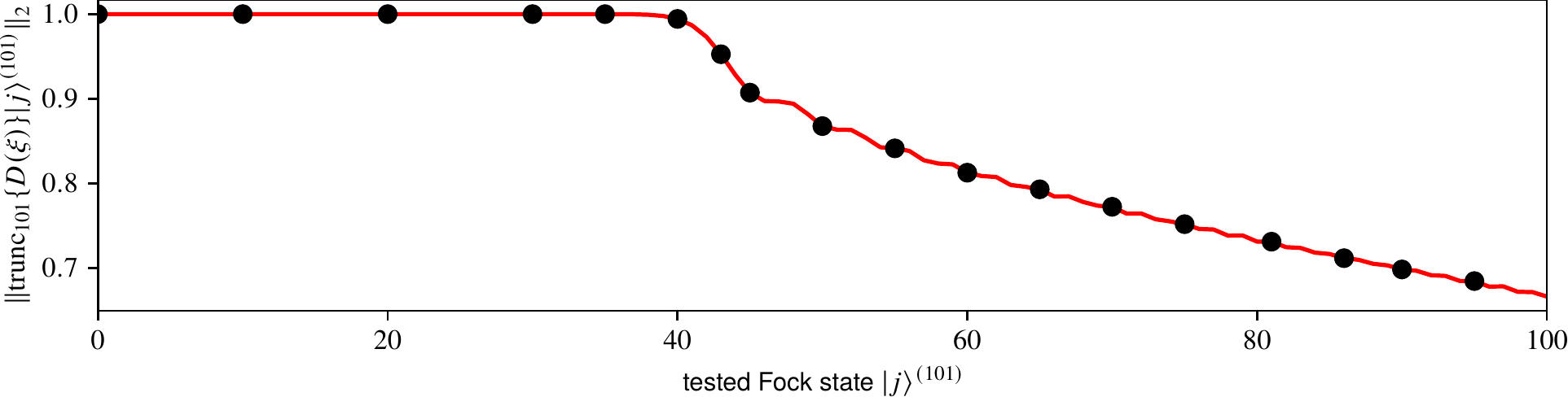}
  \caption{
    Normalisation $\norm{ \trunc_{101} \{ \hat{D}(3 - 2\imath) \} \ket{j}^{(101)} }_{2}$ of displaced Fock states $0 \leq j \leq 100$. The matrix was constructed using the closed form formula (black bullets) and approximated with TAME (red, solid) where we set $d_{0} = 101$ and $d_{1} = 161$. In both plots the dimension~$d_{1}$ for TAME was determined via Algorithm~\ref{alg:fastd1search}.
  }
  \label{dtexpmt}
\end{figure}

In Fig.~\ref{dtexpmt} we compare $\trunc_{101} \{ \hat{D}(3 - 2\imath) \}$ constructed using the closed form formula~\eqref{cahill1969} and approximated with TAME. We chose the dimension $d_{0}$ and the displacement magnitude $\abs{\xi}$ to accommodate the limits established in Section S1 of the Supplementary material. The secondary dimension $d_{1} = 161$ was chosen high enough to suppress the effects of truncation errors. The plot suggests that our method produces results equal to the closed form formula in terms of the normalisation \eqref{dxi-norm}. Further comparison of individual matrix elements reveals that, on average, the approximate matrix matches \eqref{cahill1969} up to $14$ decimal places with the worst difference matching only up to $11$ decimal places.

What remains to be determined is the proper choice, or rather the methodology of choosing a sufficiently large working dimension $d_1$ given the target dimension $d_0$. In the subsequent paragraphs we are going to show that it is practical to set the dimension $d_{1}$ as small as possible. The de facto standard \textbf{scaling and squaring} matrix exponentiation algorithm~\cite{higham2005, almohy2010} relies on matrix multiplication with the actual number of matrix products depending on the binary logarithm of the $1$--norm~\cite{golub2013} of the exponentiated matrix.

The $1$--norm~\cite{golub2013} of the $\trunc_{d_{1}} \{ \hat{Q} \}$ argument inside the matrix exponential within \eqref{tame} reads
\begin{equation}
  \norm{ \trunc_{d_{1}} \{ \hat{Q} \} }_{1} =
  \norm{ 
    \xi \trunc_{d_{1}} \{ \hat{A}^{\dagger} \} - 
    \conj{\xi} \trunc_{d_{1}} \{ \hat{A} \} 
  }_{1} =
  \abs{ \xi } \left( \sqrt{d_{1} - 1} + \sqrt{d_{1} - 2} \right) 
  \approx
  2 \abs{ \xi } \sqrt{d_{1}} \Qc
\end{equation}
where the final approximation holds asymptotically. Therefore the asymptotic computational complexity of the matrix exponential in \eqref{tame} scales as~$\bigO(\log_{2} d_{1})$ in terms of matrix products. The complexity of each matrix multiplication, specified in terms of FP operations, depends on the algorithm it utilizes. A naive textbook implementation scales as poorly as~$\bigO(d_{1}^{3})$, whereas the more sophisticated Strassen algorithm~\cite{strassen1969} scales approximately as~$\bigO(d_{1}^{2.807})$. Consequently the computational complexity of~\eqref{tame} scales as $\bigO(d_{1}^{2.807} \log_{2} d_{1})$ under optimal conditions. It is therefore imperative to keep the dimension $d_{1}$ as low as possible.

\begin{algorithm}[h]
  \begin{algorithmic}[1]
    \Procedure{FindDimension~}{
        $d_{0}, \epsilon_{1} \coloneqq 10^{-13}, h \coloneqq 10$
    }
      \State $q \coloneqq d_{0} + 1$
      \State $M_{q} \coloneqq \tame (Q, q, d_{0})$

      \While{$q < h \cdot d_{0}$}
        \State $p \coloneqq q + 1$
        \State $M_{p} \coloneqq \tame (Q, p, d_{0})$
        \If{ $\norm{ M_{q} - M_{p} }_{\max} < \epsilon_{1}$ }
          \State \textbf{return} $q$
        \EndIf

        \State $q \coloneqq p$
        \State $M_{q} \coloneqq M_{p}$
      \EndWhile
      \State \textbf{raise error} No solution found.
    \EndProcedure
  \end{algorithmic}
  \caption{
    A simple iterative search for the least dimension $q$ such that there is a match on $d_{0}$ dimensional region between two $q$ and $p \coloneqq q + 1$ dimensional matrices constructed using TAME. Here, the element-wise matrix max-norm is defined with~$\norm{M}_{\max} \coloneq \max_{ij} \abs{[M]_{ij}}$.
  }
  \label{alg:fastd1search}
\end{algorithm}

We propose a simple iterative algorithm for finding optimal $d_{1}$. Suppose a sufficiently sized $\expm (\trunc_{q} \{ \hat{Q} \})$ matrix is correct on some region spanning $\{ \ket{0}^{(u)}, \dotsc, \ket{u - 1}^{(u)} \}$ where $u \leq q$. Suppose the matrix exponential ($\expm$) algorithm is also consistent: for a differently sized $\trunc_{p} \{ \hat{Q} \}$ matrix with dimension $p > q$ the computed matrix exponential is correct on a region of at least the same size.
Given these assumptions, which are upheld by the standard $\expm$ implementation \cite{higham2005, almohy2010}, we introduce the Algorithm~\ref{alg:fastd1search} as follows. First we take the desired dimension $d_{0}$ of the correct region and set an equality tolerance $\epsilon_{1}$ for small numbers: our condition with $\epsilon_{1} = 10^{-13}$ proclaims two numbers identical if they match up to their twelfth decimal place. Then we search for a pair of larger matrices such that their $d_{0}$ regions match. The search process is significantly simplified by taking the dimension of the second larger matrix to be constantly shifted from the first larger matrix. To improve its speed we always recycle one of the matrices in the next iteration instead of recalculating it every time. The depth of the search is specified by the factor $h$. In our experience the dimension is found somewhere well below~$q = 3 \cdot d_{0}$ in the case of displacement, hence we set the depth $h$ above that. Once the search algorithm finishes \textit{successfully}, we obtain $d_{1}$.


\section{Verification of approximated matrices}

In general, we can not verify the matrix~\eqref{tame} constructed via TAME simply by comparing its elements against some exact solution for the obvious reason: if we knew the exact solution we would not be in this situation in the first place.

We have used normalisation \eqref{dxi-norm}, or more precisely the implied necessary condition of unitarity $\max_{ij} \abs{ [ G ]_{ij} } \leq 1$, to detect outright incorrect matrices in Fig.~\ref{cond2}, but alas, necessary conditions alone can not be used to prove the matrix correct. In Fig.~\ref{cond2} we determined that employing the recurrent formula \cite{miatto2020} in construction of $\trunc_{\ff} \{ \hat{D}(\xi) \}$ was ill-advised due to accumulation and consequent amplification of rounding errors over the course of the computation. While we can not safely use the recurrent formula to construct an arbitrary truncated displacement matrix, we can use it to determine whether a candidate matrix, for example one constructed via TAME \eqref{tame}, possesses appropriate structure as the formulae define relations between neighboring matrix elements.

We can repurpose the relations Eq.~(56--58) from \cite{miatto2020} to construct an error matrix
\begin{equation}\label{errormatrix}
  \begin{aligned}
    [E]_{0, 0} & =
      [G]_{0, 0} - \exp\left( - \frac{1}{2} \abs{\xi}^{2} \right)
    \\
    [E]_{i, 0} & =
      [G]_{i, 0} -
      \frac{\xi}{\sqrt{i}} [G]_{i - 1, 0}
    \\
    [E]_{i, j} & = 
      [G]_{i, j} -
      \left(
        \frac{\sqrt{i}}{\sqrt{j}} [G]_{i - 1, j - 1} -
        \frac{\conj{\xi}}{\sqrt{j}} [G]_{i, j - 1}
      \right)
  \end{aligned}
\end{equation}
for a given candidate matrix $G$. The rounding errors are not amplified in computation of the error matrix as there is no recursion. Its elements $[E]_{ij}$ give the difference between the actual elements $[G]_{ij}$ of the candidate matrix and the values they should have been based on their neighbors, $[G]_{i - 1, j - 1}$ and $[G]_{i, j - 1}$, and the structural constraints given in~\cite{miatto2020}.

\begin{figure}[h]
  \includegraphics[width = \columnwidth]{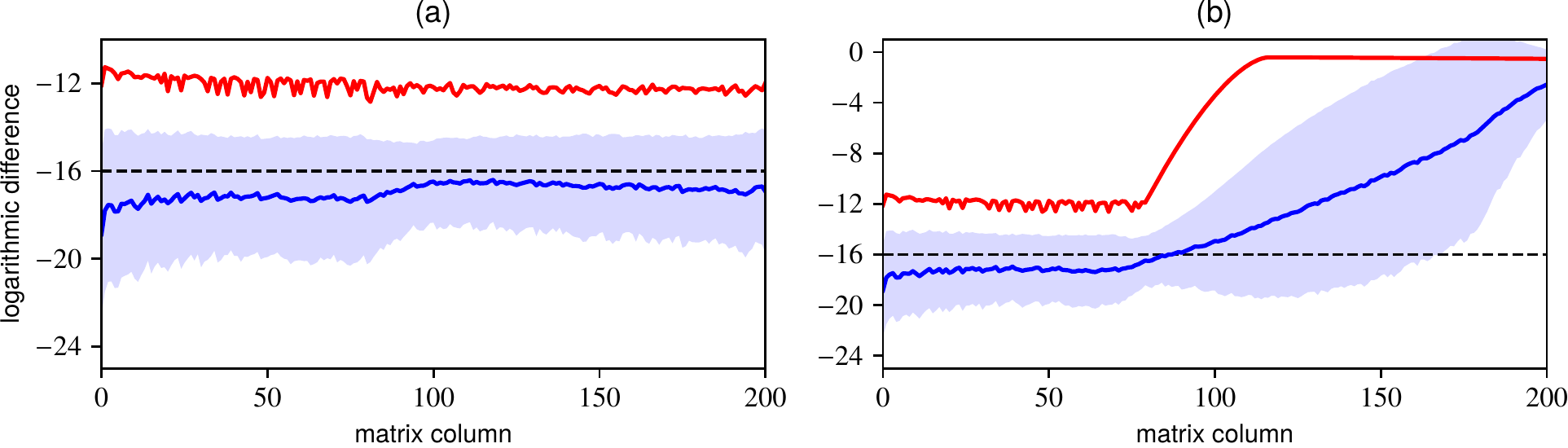}
  \caption{
    Verification of the $\trunc_{201} \{\hat{D}(3 - 2\imath)\}$ matrix approximated using (a) TAME ($d_{0} = 201$ and $d_{1} = 277$) and (b) plain matrix exponential ($d_{0} = 201$).
    Blue lines mark the row-wise $\mean_{i}(L_{ij})$ values, light-blue region stretches a standard deviation $\std_{i}(L_{ij})$ away from the mean. The maximal difference $\max_{i}(L_{ij})$ within each column is represented by the red line. The dashed horizontal line corresponds to the unit round-off in double precision floating point number representation.
    \textbf{(a)}~The matrix is structurally correct. The average differences are negligible, their values falling below the unit round-off. The maximal differences match up to $11$ decimal places. 
    \textbf{(b)}~The matrix maintains correct structure in its first third.  The truncation errors manifest in the rest of the matrix as an exponential explosion in the maximal difference (around the 100th column) and a steady rise in the mean value. 
  }
  \label{error-3}
\end{figure}

In Fig.~\ref{error-3} we compare the decadic logarithm of the difference ${[L]_{ij} = \log_{10} \abs{ [E]_{ij} }}$ for ${\trunc_{201} \{ \hat{D}(3 - 2\imath) \}}$ approximated using (a) TAME (${d_{0} = 201}$, ${d_{1} = 277}$) and (b) the plain matrix exponential~(${d_{0} = 201}$). In each plot we display the row-wise $\mean_{i}([L]_{ij})$ value with blue line. The surrounding light-blue area stretches one standard deviation $\std_{i}([L]_{ij})$ from the mean. The maximal difference $\max_{i}([L]_{ij})$ within each column is represented by the red line. Finally the dashed black horizontal line (at $-16$) roughly corresponds to the double precision unit round-off~\cite{higham2002}.

In Fig.~\ref{error-3}~(a) the matrix is structurally correct, with the maximal difference still matching up to $11$ decimal places. On average the differences fall below the unit round-off, essentially making the errors negligible. In Fig.~\ref{error-3}~(b) the matrix constructed using the plain matrix exponential maintains the correct structure in the first third of its columns, however, the truncation errors begin to manifest at that point. This can be observed as an exponential explosion in the maximal difference (around the 75th column) and a steady rise in the mean value. We saw a similar manifestation of truncation errors in Fig.~\ref{cond2} where the columns incorrectly retained their normalization as if the truncated matrix remained unitary.

\section{Numerical simulation of the preparation circuit}
\label{simul}

The CV nature of the preparation scheme in Fig.~\ref{scheme}, described with relations \eqref{schR} and \eqref{schP}, makes its exact numerical simulation not only impractical, but outright impossible. We can, however, perform an approximate numerical simulation of the formulae on a TFS. We have already proposed TAME as the method for approximating the truncated displacement operator. We have yet to ascertain a key ingredient of the simulation. We must determine the optimal dimension $d_{0}$ of the TFS, which should be large enough to support all the quantum states occurring in the simulation. 

Following the Fig.~\ref{scheme}, we begin with the TMSV state. One of its modes is attenuated by the $\mathcal{G}_{2}^{\eta}$ channel. This only reduces its energy and, as a consequence, the required support shrinks in size. We can therefore safely disregard the attenuating channel and simplify the expression for the marginal state \eqref{schR} into ${ \hat{\varrho} \propto \trx_{2} \left\{ \hat{D}_{2}(\xi) \ketbra{\gamma}[1, 2]{\gamma} \hat{D}_{2}(\xi)^{\dagger} \hat{\Pi}_{2} \right\} }$. We then require that both the initial and the displaced TMSV states are faithfully approximated on the $d_{0}$ dimensional TFS for \textit{all the possible} values of $\gamma$ and $\xi$. By taking the largest displacement $\xi^{\star}$ and squeezing rate $\gamma^{\star}$ considered in the simulation, we can iteratively determine $d_{0}$ as the least dimension such that the cutoff error~\eqref{error} falls below some threshold $\epsilon_{0}$. This condition reads
\begin{equation}
  1 - \sum_{i = 0}^{d_{0} - 1} \sum_{j = 0}^{d_{0} - 1} \Abs{ [ \ket{\gamma^{\star}} ]^{S}_{ii} [ \hat{D} (\xi^{\star}) ]^{S}_{ij} }^{2} \leq \epsilon_{0}
\end{equation}
for the displaced TMSV state. While the coefficients ${ [ \ket{\gamma^{\star}} ]^{S}_{ii} = \cosh^{-1} \gamma^{\star} \tanh^{i} \gamma^{\star} }$ of the TMSV state are determined trivially, the matrix elements $[ \hat{D} (\xi^{\star}) ]^{S}_{ij}$ of the displacement operator can not be, in general, obtained analytically with \eqref{cahill1969} and we must employ alternate means such as TAME.

In the simulation we consider $0 \leq \gamma \leq 1$, corresponding to roughly $8.7 \dB$ squeezing, and $0 \leq \xi \leq 1$, hence we set $\gamma^{\star} \equiv \xi^{\star} \equiv 1$ while searching for $d_{0}$. Once the dimension $d_{0}$ is found, we determine its respective $d_{1}$ using the Algorithm~\ref{alg:fastd1search}. With the thresholds $\epsilon_{0} \equiv \epsilon_{1} \equiv 10^{-13}$ we get $d_{0} = 70$ and $d_{1} = 90$ for $\gamma^{\star} \equiv \xi^{\star} \equiv 1$. Note that for this particular $d_{1}$, the TAME matrices constructed on $d_{1}$ and $d_{1} + 1$ dimensional TFS are identical in double precision FP arithmetic.

In the following sections we use the numerical methodology we developed to determine the benefits of using PNRD, APD, and APD cascades in a pair of applications of the preparation circuit. First we discuss preparation of non-linearly squeezed states (Section~\ref{sec-nlss}) and then follow with construction of well defined non-classical superpositions of Fock states (Section~\ref{sec-fock}). In both applications the figures of merit are functions depending on the resulting density matrix \eqref{schR} and the associated probability of success \eqref{schP}. We approach the analysis with a rudimentary grid based exploratory strategy for optimization. We divide the $[ 0 \leq \gamma \leq 1 ] \otimes [ 0 \leq \xi \leq 1] $ region into equidistant $1001 \times 1001$ grid of points ${ p_{j} \coloneqq (\gamma_{j}, \xi_{j}) }$ and evaluate the numerically approximated relations~\eqref{schR} and~\eqref{schP} for each point $p_{j}$ and each experimental scenario ${ q_{i} \coloneqq (\eta_{i}, \pi_{i}) }$ defined by the overall efficiency ${ \eta_{i} \in \{ 0.80, 1.00 \} }$ of the setup and expected measurement outcome $\pi_{i}$ respective to the POVM elements. These entail
\begin{equation}\label{pin}
  \eye_{2} - \ketbra{0}[2]{0}, 
  \ketbra{1}[2]{1}, 
  \ketbra{2}[2]{2}, 
  \ketbra{3}[2]{3}, 
  \ketbra{4}[2]{4}, 
  \ketbra{5}[2]{5}, 
  \ketbra{6}[2]{6}
\end{equation}
representing the click of the ideal APD and the first six PNRD outcomes relevant in our preparation scheme, as well as  
PNRD approximations employing APD cascades
\begin{equation}\label{piMn}
  \hat{\Pi}_{1}^{10}, \hat{\Pi}_{1}^{5}, \hat{\Pi}_{1}^{2},
  \hat{\Pi}_{3}^{10}, \hat{\Pi}_{3}^{5}, \hat{\Pi}_{3}^{4}
\end{equation}
where $\hat{\Pi}_{n}^{M} \coloneqq \sum_{k = 0}^{d_{0} - 1} p_{M} (n|k) \ketbra{k}[2]{k}$. 
The POVM elements $\hat{\Pi}_{n}^{M}$ represent outcomes where exactly $n$ detectors click within APD cascade comprising $M$ detectors~\cite{provaznik2020}. The individual probabilities $p_{M}(n|k)$ read
\begin{equation}
  p_{M} (n|k) \coloneqq M^{-k} \sum_{l = 0}^{n} \binom{n}{l} (-1)^{l} (n - l)^{k} \Qd
\end{equation} 
This way we procure an assortment of probabilities $P (i, j)$ and density matrices $\varrho (i, j)$ corresponding to the $p_{i}$ and $q_{j}$ sequences. We then utilize these values in objective and constraint functions, that will be discussed in detail in the following sections, to analyze the performance of the preparation scheme in particular applications and its response to different experimental configurations.

The numerical simulation and the analysis of its results was implemented using a number of open source software libraries~\cite{harris2020,virtanen2020,johansson2013,dalcin2021,hunter2007,jupyter,sympy} in Python. 


\section{Engineering non-linearly squeezed states}
\label{sec-nlss}

Nonlinear squeezing was originally introduced~\cite{miyata2016} as a measure quantifying the quality of approximate cubic states suitable for optical measurement-induced quantum gates~\cite{miyata2016,marek2018}. It has been shown to apply to higher ordered phase squeezing gates as well~\cite{marek2018} and was recently discussed in detail~\cite{kala2021}.
The ideal cubic operation facilitates unitary evolution with interaction Hamiltonian proportional to~$\hat{X}^3$. When approximatively implemented in the measurement induced fashion~\cite{miyata2016} its action in the Heisenberg picture can be represented by the operator transformation
%
%
\begin{equation}
  \begin{aligned}
    \hat{X}_{S} & \to 
      \hat{X}_{S} \Qc \\
    \hat{P}_{S} & \to 
      (\hat{P}_{S} + \nu \hat{X}_{S}^{2}) +
      (\hat{P}_{A} - \nu \hat{X}_{A}^{2}) \Qc
  \end{aligned}
\end{equation}
where the $\hat{X}_{S}, \hat{P}_{S}$ operators correspond to the signal state and $\hat{X}_{A}, \hat{P}_{A}$ to some ancillary mode. The first terms of both relations correspond to the ideal cubic interaction $\exp(\imath \frac{\nu}{3} \hat{X}^{3})$. 
The additional term (${\hat{P}_{A} - \nu \hat{X}_{A}^{2}}$) represents the \emph{nonlinear quadrature} of the ancillary mode and embodies the undesirable noisy contribution. It can be suppressed by choosing an appropriate ancillary state with the right structure. Effects of this contribution, or more precisely its variance and mean, vanish for the ideal cubic state. In general, neither the variance nor the mean vanish for physical approximations of the ideal cubic state. Good approximations, however, minimize their values and consequently the variance of this contribution may be used to quantify the quality of these approximate cubic states.

The preparation scheme presented in Fig.~\ref{scheme} can be utilized for production of quantum states approximating the ideal cubic state. We have discussed the methodology of the simulation in detail in Section~\ref{simul}. In essence we search for optimal values of squeezing $\gamma$ and displacement $\xi$ that lead to high quality cubic state approximations while maximizing the probability of successful preparation. The optimization is performed for various experimental scenarios involving different detectors and taking a range of overall losses into account. 

To measure the approximation quality we adapted the nonlinear quadrature and the concept of nonlinear squeezing discussed in \cite{miyata2016} to fit our simulation.
We employ the nonlinear quadrature $\hat{Y} = \mu \hat{P} - \sqrt{2^{-1}} \mu^{-2} \hat{X}^{2}$ and use its variance ${ V(\hatrho) = \expvalue{ (\hat{Y} - \expvalue{\hat{Y}}_{\hatrho})^{2} }_{\hatrho} }$ to measure the nonlinear squeezing of arbitrary states~$\hatrho$.
Potential effects of Gaussian squeezing \cite{miyata2016,kala2021} on $V(\hatrho)$ are eliminated by minimizing over 
the parameter $\mu$. Consequently we base our analysis on the minimized quantity ${M (\hatrho) \coloneqq \lambda_{G}^{-1} \min_{\mu} V(\hatrho)}$ normalized with respect to the minimal variance $\lambda_{G} \coloneqq \min_{\hatrho_{G}} \min_{\mu} V(\hatrho_{G}) \equiv 0.75$ achievable by Gaussian states $\hatrho_{G}$~\cite{kala2021}. 

The numerical simulation yields density matrices $\varrho(i, j)$ along with the $P(i, j)$ probabilities of success corresponding to different experimental parameters. We then compute the individual moments required in the calculation of $V(\hatrho)$ from the elements of density matrices $\varrho(i, j)$. We avoid the matrix representation of the operators in the computation to avert truncation errors and employ closed form formulae instead. The minimization with respect to $\mu$ within $M(\hatrho)$ is solved analytically. 

We thus obtain $M(i, j)$ values for their respective $\varrho (i, j)$ matrices and $P (i, j)$ probabilities. We then divide the dataset corresponding to each experimental scenario $q_{i}$ into bins based on values of the variance $M(i, j)$ and find the maximal attainable probability $P(i, j)$ within each bin.

\begin{figure}[h]
  \includegraphics[width = \columnwidth]{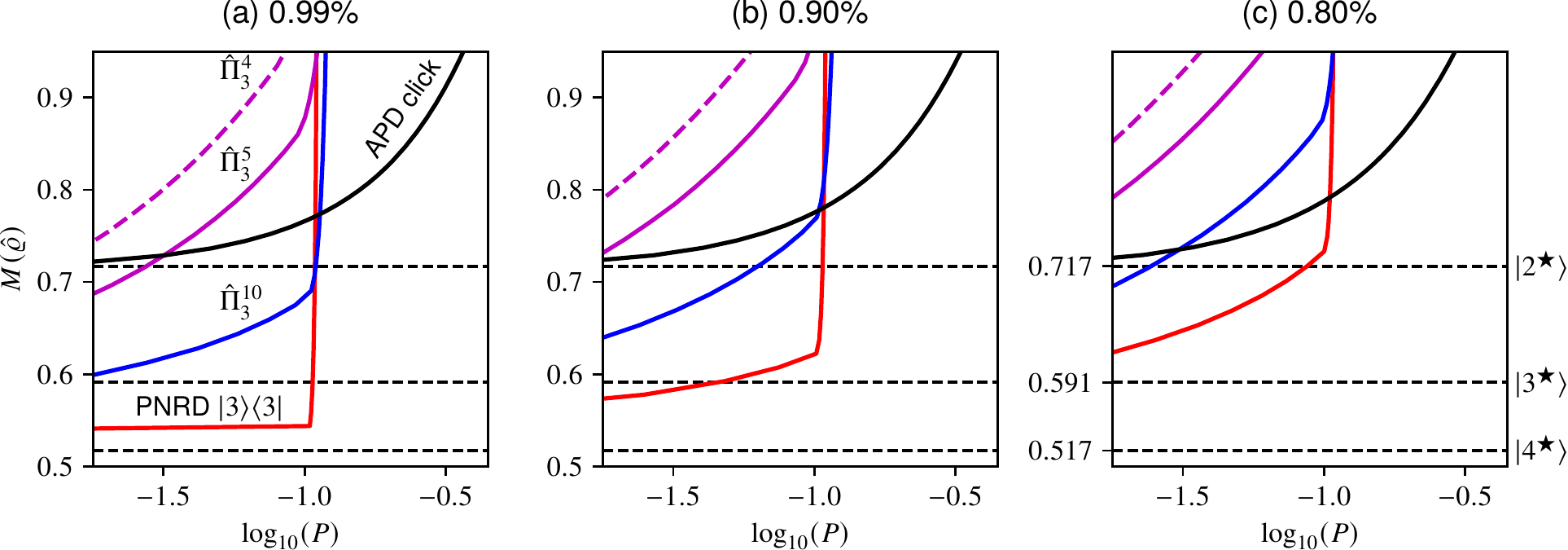}
  \caption{
    A comparison of attainable variance $M(\hatrho)$ as a function of success probability. The variance is normalized with respect to the minimal variance achievable by Gaussian states. We use the same vertical and horizontal axes in the plots to show the contrast between the almost ideal (a) and lossy (b, c) scenarios with $99\%$, $90\%$ and $80\%$ transmission efficiencies. Horizontal dashed lines are used to mark the optimal cubic state approximations $\ket{v^{\bigstar}} \in \sH_{v}$ constructed on low-dimensional TFS. We encode the information about the POVM elements as follows: APD click with solid black line, PNRD projection onto $\ket{3}$ with solid red, APD cascades comprising four ($\hat{\Pi}_{3}^{4}$, dashed magenta), five ($\hat{\Pi}_{3}^{5}$, magenta) and ten ($\hat{\Pi}_{3}^{10}$, blue) detectors where three detectors click. Overall, utilizing the PNRD $\ket{3}$ (solid red) produces states with lowest non-linear variance, therefore producing comparatively better approximations of the cubic state. In both (b) and (c) a single APD outperforms the APD cascades comprising five and four detectors for probabilities greater than~$1\%$. In this regime the cascade comprising ten detectors still offers advantage over single APD. In (c) a single APD outperforms APD cascades comprising either four, five or ten detectors for success probabilities larger than roughly~$5\%$.
  }
  \label{VP}
\end{figure}

In Fig.~\ref{VP} we present a comparison of the attainable variance $M(\hatrho)$ as a function of success probability $P$. We examine different detection outcomes, in particular the PNRD projection onto $\ket{3}$ (red line) and its three approximations realized through an APD cascade \cite{provaznik2020} where three APD detectors out of four (dashed magenta), five (magenta) and ten (blue) click. Their respective POVM elements $\hat{\Pi}_{3}^{4}$, $\hat{\Pi}_{3}^{5}$ and $\hat{\Pi}_{3}^{10}$ are given by the relation \eqref{piMn}. 
We consider a single APD detector (black line) as well. The plots show (a) $99\%$, (b) $90\%$, and (c) $80\%$ transmission efficiency $\eta$. The results are normalized with respect to the minimal variance  achievable by Gaussian states. 
The optimal cubic state approximations~\cite{miyata2016} $\ket{v^{\bigstar}} \in \sH_{v}$ constructed on $v$ dimensional TFS are marked with dashed horizontal lines.
These states were found by searching for pure states spanning the first $v$ Fock states that would minimize the variance $M(\hatrho)$ of the non-linear quadrature~\cite{miyata2016}.
Their inclusion makes it possible for qualitative comparison with the states produced by our scheme.

In general using PNRD yields the best results. In the idealized scenario with $99\%$ efficiency the PNRD projecting onto~$\ket{3}$ approaches the variance of the optimal $\sH_{4}$ non-linearly squeezed state $\ket{4^{\bigstar}}$. It also attains the best values consistently across the considered transmission efficiencies, therefore producing comparatively better approximations of the cubic state than either the APD cascades or a single APD. 
In the $90\%$ and $99\%$ regimes, the APD cascade comprising ten detectors promises better performance than a single APD or any other cascade configuration for that matter.
In the low-efficiency mode ($80\%$) we can see that a single APD outperforms APD cascades for  probabilities of success greater than $5\%$. 
This can be attributed to the imperfections inherent to APD cascades~\cite{provaznik2020}. Their flaws become emphasized with increased loss, rendering a single APD to be the better choice.

In conclusion, unless a PNRD capable of distinguishing at least three photons is available, it is advantageous to use a single APD in any practical scenario with non-ideal transmission efficiency as long as success probabilities larger than approximately $5\%$ are desired. The advantage of a single APD can be offset by using an exorbitant number of detectors within APD cascade. 


\section{Preparation of high fidelity qubit in Fock basis}
\label{sec-fock}

The qubit-like superposition $\ket{\vartheta} \coloneqq \cos \vartheta \ket{0} + \sin \vartheta \ket{1}$ represents one of the simplest non-Gaussian quantum states of light. 
It serves an important role in quantum information processing~\cite{kok2010} and is one of the resources available in contemporary experimental quantum optics\cite{davidson2021,higginbottom2016,podhora2017}. As such it is has been employed in experimental demonstrations of various theoretical concepts including witnessing of non-Gaussianity~\cite{filip2011,jezek2011,straka2014,mika2022} and hybrid entanglement~\cite{huang2016,lejeannic2018} in quantum communication.

This family of quantum states can be produced with the preparation scheme we have previously introduced in Fig.~\ref{scheme}. We can search for the optimal squeezing $\gamma$ and displacement $\xi$ parameters to obtain a given target state $\ket{\vartheta}$ with sufficient fidelity $F = \braket{\vartheta | \hatrho | \vartheta}$ and maximal performance in terms of success probability.

We compute the corresponding $F_{\vartheta} (i, j)$ values for the $P (i, j)$ probabilities and $\varrho (i, j)$ density matrices obtained in the simulation described in detail in Section~\ref{simul}. We then divide the dataset for each experimental scenario $q_{i}$ into bins comprising subsets of data satisfying $F_{\vartheta} (i, j) \geq \tau$ where $\tau$ specifies a moving fidelity threshold. The maximal attainable probability $P(i, j)$ is then found for each subset.

\begin{figure}[h]
  \includegraphics[width = \columnwidth]{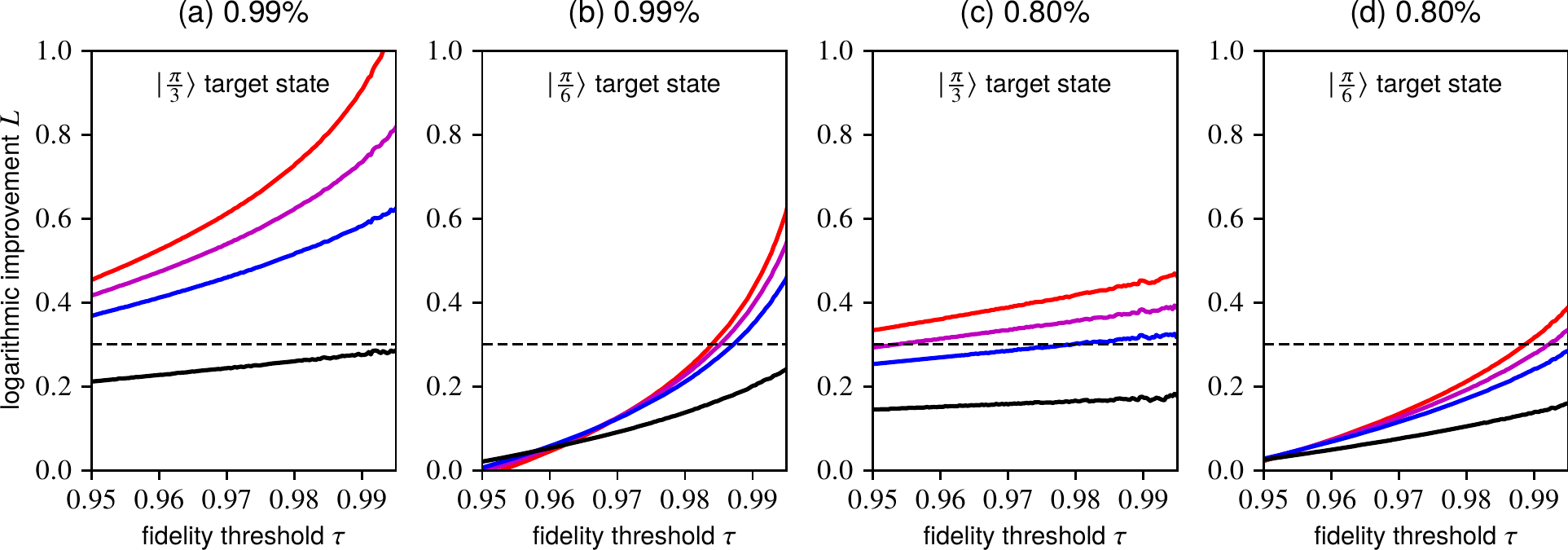}
  \caption{
    Benchmarking the performance of PNRD and its approximations using APD cascades relative to a single APD detector in preparation of particular superpositions ${\ket{\vartheta} \coloneqq \cos \vartheta \ket{0} + \sin \vartheta \ket{1}}$ parametrized with $\vartheta \in \reals$. The PNRD projection onto $\ket{1}$ is represented by red line, whereas the magenta line corresponds to APD cascade comprising ten detectors where a single detector clicks ($\Pi_{1}^{10}$), blue line to cascade of five detectors ($\Pi_{1}^{5}$) and black line depicts the case with two detectors~($\Pi_{1}^{2}$). The plots demonstrate preparation of two distinct states while considering different transmission efficiencies. In (a) and (c) we aim to prepare $\ket{\frac{\pi}{3}}$. In (b) and (d) we target $\ket{\frac{\pi}{6}}$. In plots (a) and (b) we consider $99\%$ transmission efficiency, while in (c) and (d) we consider mere $80\%$. The horizontal dashed line marks a twofold improvement in each plot.
    \textbf{(a)} In the high-efficiency regime we obtain roughly tenfold improvement in the high-fidelity preparation of the $\ket{\frac{\pi}{3}}$ state. The PNRD approximations using more than two detectors offer a significant improvement as well.
    \textbf{(b)} While the advantage of PNRD is reduced when targeting the state biased towards $\ket{0}$, it still offers roughly four times better performance.
    \textbf{(c)} The PNRD detector and its approximations offer 2-3x higher success probability even in the lower-efficiency scenario. 
    \textbf{(d)} The PNRD detector and its approximations offer roughly twofold improvement.
  }
  \label{fp-1}
\end{figure}

In Fig.~\ref{fp-1} we demonstrate the relative improvement in probability of successfully engineering $\ket{\vartheta}$ states by employing different detectors instead of a single APD. We consider a pair of target states, $\ket{\frac{\pi}{3}}$ and $\ket{\frac{\pi}{6}}$, both evaluated for $99\%$ and $80\%$ transmission efficiencies. These target states were chosen to probe the improvement for unbalanced superpositions biased either towards $\ket{0}$ or $\ket{1}$ states. In the plot we show the result obtained for projection onto $\ket{1}$ (red line) realized by PNRD and the results obtained with its approximations realized through APD cascades where a \textit{single} detector out of ten ($\hat{\Pi}_{1}^{10}$, magenta), five ($\hat{\Pi}_{1}^{5}$, blue) and two ($\hat{\Pi}_{1}^{2}$, black) clicks. The POVM elements $\hat{\Pi}_{n}^{M}$ of the cascades were defined in \eqref{piMn}. The figure of merit is defined as $L \coloneqq (\log_{10} P_{\bullet} - \log_{10} P_{\textrm{APD}})$ with $P_{\bullet}$ respective to individual detection outcomes.

In general, conditioning on the PNRD $\ket{1}$ detection outcome yields the best results. In the high-fidelity regime with $99\%$ efficiency the relative improvement is roughly tenfold for $\ket{\frac{\pi}{3}}$ and roughly four times better for $\ket{\frac{\pi}{6}}$. The APD cascades comprising ten and five detectors follow. The relative lead of the PNRD diminishes in the $80\%$ efficiency regime. Its advantage also dwindles when we consider target states closer to $\ket{0}$, such as the $\ket{\frac{\pi}{6}}$ state. While the cascade comprising two detectors falls short in every case, it still outperforms a single APD, albeit not by a lot.


\section{Conclusions}


We have analyzed the numerical accuracy of several currently available methods~\cite{miatto2020,cahill1969} used in construction of the truncated coherent displacement operator, an essential ingredient of state preparation in quantum optics~\cite{dakna1999,gottesman2001,fiurasek2005,marek2011} and many protocols used in quantum information processing~\cite{adesso2014,braunstein2005,gottesman2001,cahill1969,weedbrook2012,lloyd1999}. We have proposed an alternative approach promising a better accuracy. Our method is based on the standard matrix exponential~\cite{higham2005,almohy2010} with truncated argument. We compute the matrix exponential on a higher-dimensional space and truncate the resulting matrix to the target dimension, thus stripping erroneous matrix elements away from the truncated displacement operator. To avoid negatively impacting computational performance, the higher dimension should be ideally kept as low as possible. To this end we provide an off-line search algorithm that can be used to determine its optimal value. To ascertain the accuracy of the resulting matrix we complement the construction method with a verification strategy based on the recurrent formulae discussed in~\cite{miatto2020}.

We have used our construction method for analysis of non-Gaussian state preparation scheme based on suitable manipulation of a two mode squeezed vacuum with subsequent photon counting measurement~\cite{gottesman2001,yukawa2013,yukawa2013b,bohmann2018} in the context of engineering non-linearly squeezed cubic states~\cite{gottesman2001,ghose2007,miyata2016} for measurement induced cubic gates~\cite{miyata2016,marek2018,gottesman2001} and construction of qubit-like superpositions between vacuum and single photon states. The latter application can be verified experimentally with currently available technology. 
We have compared the effects of different detection mechanisms, including APD, PNRD and its approximations using APD cascades~\cite{provaznik2020} with varying number of APD detectors, to determine practical approach towards state preparation.
In our analysis we have optimized the free parameters of the prepraration scheme, the initial squeezing and the displacement, to attain optimal results in both applications.
This analysis also provides additional metric which can be used to quantify the quality of APD cascades. We have found that in practical applications when PNRD is not available, using a single APD to engineer non-linearly squeezed states offers better performance compared to employing APD cascades comprising small number of detectors. We attribute this counter-intuitive result to the imperfections inherent to APD cascades~\cite{provaznik2020} which are exaggerated with increased loss; these flaws became significant for $20\%$ overall loss. 
The primary cause of this behaviour lies within the employed avalanche detectors as a single click may be triggered by multiple photons. While this is a critical issue for multi-photon state engineering, it is not as significant for single-photon states. 
We have determined that using APD cascade, even if one comprising only a pair of APD detectors, improves upon using a single APD in preparation of high-fidelity non-Gaussian qubit-like superpositions.

Our circuit variant can be extended to utilize multiple displacements and detectors. Similarly the proposed method for numerical construction of truncated unitary operators is not limited to displacement only can be applied to, for example, squeezing or cubic phase-shift operators. Furthermore, the method could be employed in preparation of a wider variety of quantum states with practical applications, such as GKP states~\cite{gottesman2001}. 


\FloatBarrier


\bibliography{arxiv-revised.bib}


\section*{Funding}


We acknowledge Grant No.~GA22-08772S of the Czech Science Foundation and also
support by national funding from MEYS and the European Union’s Horizon 2020
(20142020) research and innovation framework programme under grant agreement
program under Grant No.~731473 (project 8C20002 ShoQC). Project ShoQC has
received funding from the QuantERA ERA-NET Cofund in Quantum Technologies
implemented within the European Union’s Horizon 2020 program. We have further
been supported by the European Union’s 2020 research and innovation programme
(CSA --- Coordination and support action, H2020-WIDESPREAD-2020-5) under grant
agreement No.~951737 (NONGAUSS). J. P. acknowledges project IGA-PrF-2022-005 of
Palacky University Olomouc. R.F. also acknowledges LTAUSA19099 of the Czech
Ministry of Education, Youth and Sports (MEYS).

\section*{Acknowledgements}

J.~P. acknowledges using the computational cluster at the Department of Optics for parallel execution of the numerical simulation.

\section*{Author contributions statement}

J.~P. conceived and developed the numerical techniques, performed the calculations, implemented and consequently conducted the numerical simulation. P.~M. and R.~F. conceived and discussed the applications and evaluations that were simulated. All authors analyzed and discussed the results, and contributed to the writing of the manuscript.

\section*{Competing interests}
Te authors declare no competing interests.

\section*{Data availability}
The datasets generated and analyzed in the current study are available from the corresponding author on reasonable request.

\section*{Source code availability}
The source code used to generate and analyze the datasets is available from the corresponding author on reasonable request.

\section*{Additional information}
Supplementary information is available as a separate document.

\end{document}